\documentclass{PoS}

\usepackage{amsmath}
\usepackage{amsfonts}

\newcommand{\overbar}[1]{\mkern 1.5mu\overline{\mkern-1.5mu#1\mkern-1.5mu}\mkern 1.5mu}
\def\bar{\overbar}

\title{Antiproton-proton interaction and related hadron physics}

\ShortTitle{$\overbar pp$ interaction and related hadron physics}

\author{\speaker{Xian-Wei Kang}\\
        Institute for Advanced Simulation, J\"ulich
Center for Hadron Physics, and Institut f\"ur
Kernphysik, Forschungszentrum J\"ulich, D-52425 J\"ulich, Germany\\
        E-mail: \email{x.kang@fz-juelich.de}}

\abstract{Antinucleon-nucleon interaction has been established in chiral effective field theory. The strong threshold enhancement observed in 
the reactions $J/\psi\to\gamma \overbar pp$ and $e^+e^-\to\overbar pp$ are interpreted by the strong $\overbar pp$ interaction. 
Concerning the channel $J/\psi\to\gamma \overbar pp$, the topic on the $\overbar pp$ bound state is also discussed.}

\FullConference{The 8th International Workshop on Chiral Dynamics, CD2015 ***\\
        29 June 2015 - 03 July 2015\\
        Pisa,Italy}

\begin{document}

\section{Introduction}
Few-body hadron-hadron interaction has been and is still a fundemantal constituent part of hadron and nuclear physics. Among them, the antinucleon-nucleon ($\overline NN$)
has been achieved fruitful progress, especially in a meson-exchange model, see e.g., a review in Ref.~\cite{Klempt}. And after 1990s, chiral effective field theory 
(EFT) has become a powerful tool to analyze nucleon-nucleon interaction, for a review, see e.g., Ref.~\cite{EHM}. However, only little work of $\overline NN$ in chiral EFT 
has been done, besides the one of partial-wave analysis (PWA) \cite{Zhou}. The recently resurgent interest of $\bar pp$ physics is triggered by the threshold enhancement
in $\bar pp$ invariant mass spectrum observed in experiments, e.g, for the decays $J/\psi\to\gamma \bar pp$ \cite{BES2003,BES2012}, 
$e^+e^-\to\bar pp$ \cite{BaBar2006,BaBar2013}. We will elaborate below how we use the $\bar pp$ final-state interaction (FSI) to interpret these phenomenons.

\section{$\overline N N$ interaction in chiral EFT}\label{sec:NNbar}
\begin{figure}
\centering
\includegraphics[scale=0.4,clip=true]{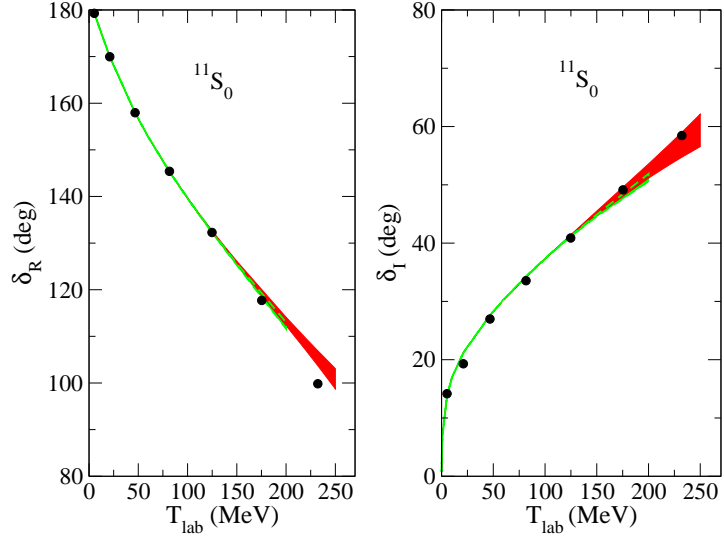}
\caption{Complex phase shifts for isospin-0 $^1S_0$\, $\left(^{11}S_0\right)$ partial wave in unit of degree as a function of kinetic laboratory
energy $T_{\text{lab}}$. $\delta_R$ coincides with the widely used conventional real phase shift while $\delta_I=-\log(\eta)/2$. 
Circle points represent PWA reported from Ref.~\cite{Zhou}. Green (Red) band indicates the cutoff dependence of NLO (NNLO) potential.}
\label{fig:11S0}
\end{figure}

$\overline NN$ potential consists of two parts: elastic scattering part and the annihilation, which is the same feature in the framework of both the conventional meson-exchange
model and the chiral EFT. The difference is on the technical treatment. In chiral EFT, the elastic part is governed by the pion exchanges (pion as the only degree 
of freedom in chiral EFT), 
which is tied closely to the knowledge of $NN$ interaction except for the sign difference due to the $G-$parity transformation. For the power counting rule and 
details in $NN$ scattering, one refers to Refs.~\cite{EHM}. Here we only remind that 
\begin{eqnarray}
V_{1\pi}^{\overbar NN}=-V_{1\pi}^{NN}\,,\quad V_{2\pi}^{\overbar NN}=V_{2\pi}^{NN}
\end{eqnarray}
because of the $G-$parity transformation rule from $\pi NN$ vertex to $\pi\overbar N\overbar N$.
The new feature of $\overline NN$ compared to $NN$ is the existence of annihilation effect that will be parameterised in contact term (in charge of short-range interaction).
In J\"ulich model \cite{Hippchen}, the annihilation is treated as a energy-, spin-, and isospin-independent Gaussian form. Here we still follow the spirit of chiral power counting, 
taking $^1S_0$ partial wave as example. 
Starting form the most elaborated couple-channel model, we have
\begin{eqnarray}\label{V_ann}
V_{\text{ann}} = \sum_{X=2\pi,3\pi,...} V_{\overline NN \to X} G_X(z)
V_{X\to \overline NN},
\end{eqnarray}
where $X$ denotes, in principle, any possible intermediate states including $2\pi,\,3\pi,\,$ etc., and $G$ is the free Green's function. It is argued that the annihilation
does not introduce a new scale into the problem \cite{NNbar}, i.e., it can be likewise treated in the chiral expansion. Then up to next-to-next-to-leading order (NNLO), one can write
the $\overline NN\to i$ ($i$ as mesons) annihilation potential in $^1S_0$ partial wave as
\begin{equation} 
V_{\overline NN\to i}=a_i+b_ip^2,
\end{equation}
where $p$\,($p'$) is the module of three-momentum in center-of-mass system (CMS) of intital (final) $\overline NN$ states.  Picking out the imaginary part, one would get 
\begin{eqnarray}\label{ImV}
\text{Im}\,V_{\text{ann}} (^1S_0)=-\left(\tilde C^a_{^1S_0}+C^a_{^1S_0} p^2\right)\left(\tilde C^a_{^1S_0}+C^a_{^1S_0} p'^2\right).
\end{eqnarray}
Equation \eqref{ImV} fulfils the unitary condition by definition (an alternative method is based on the dispersion theory, see e.g., the $\pi\pi$ sector in Ref.~\cite{Bpipi}). 
Expanding the real part from the principal integral, we will get the similar structure as the $NN$ case,
\begin{eqnarray}
\text{Re}\,V_{\text{ann}} (^1S_0)=\tilde C_{^1S_0}+C_{^1S_0} \left(p^2+p'^2\right).
\end{eqnarray}
Note there are four LECs ($\tilde C_{^1S_0}^a,\, C_{^1S_0}^a,\,\tilde C_{^1S_0},\,C_{^1S_0}$) in total, where the annihilation in indicated in the superscript by ``a''.
Now the $\overline NN$ potential is setting up but containing the low-energy constants (LECs). The recent energy-dependent $\overline NN$ partial-wave analysis (PWA) 
is done in Ref.~\cite{Zhou},
which provides a rather nice description of all the $\bar pp$ scattering data below laboratory momentum of $925\,\text{MeV}$. 
These LECs will be fitted to the partial-wave amplitude there.
The results for the inelasticity and phase shfits of $^{11}S_0$ are shown in Fig.~\ref{fig:11S0} up to $T_{\text{lab}}=200\,\text{MeV}$ for NLO and 
$T_{\text{lab}}=250\,\text{MeV}$ for NNLO. $T_\text{lab}$ is the kinetic laboratory energy, $T_{\text{lab}}=2k^2/m$ with $k$ denoting the module of the on-shell momentum
in CMS. We have used the notation $^{2I+1\,\,2S+1}L_J$, where $L,\,S,\,J$ denote the orbital 
angular momentum, total spin and their quantum addition, i.e, total angular momentum, respectively. The phase shift (complex value due to annihilation) is 
defined from $S-$matrix as $S=\eta e^{2i\delta_R}\equiv e^{2i\delta}$ with $\delta\equiv\delta_R+i\delta_I$, and then $\delta_I=-\log(\eta)/2$. 
The band is formed by varying the cutoff combination applied into the 
Lippmann-Schwinger equation and spectral function in the two-pion exchange potential \cite{SFR,NNbar}. From Fig.~\ref{fig:11S0} one sees that our results reproduce the PWA rather well
with very small uncertainty (cutoff dependence) at the whole region considered. The corresponding results for coupled partial wave $^3S_1 - {}^3D_1$ will be used
in the following subsection but are not shown here due to the limited room. For the reader who is interested in this part, one refers to the publication in \cite{NNbar}.
Besides the phase shifts and inelasticities, the scattering length, and the level shifts and widths of the antiprotonic hydrogen are also calculated.
They are all in good agreement with experimental numbers within error bars. $\overline NN$ Bound states
are predicted in $^{13}P_0$ and $^{13}S_1 - {}^{13}D_1$ partial waves \cite{NNbar}. 

\section{$\bar pp$ related hadron physics}

\subsection{$\bar pp$-threshold enhancement in $J/\psi\to\gamma\bar pp$}
\begin{figure}
\centering
\includegraphics[scale=0.36,clip=true]{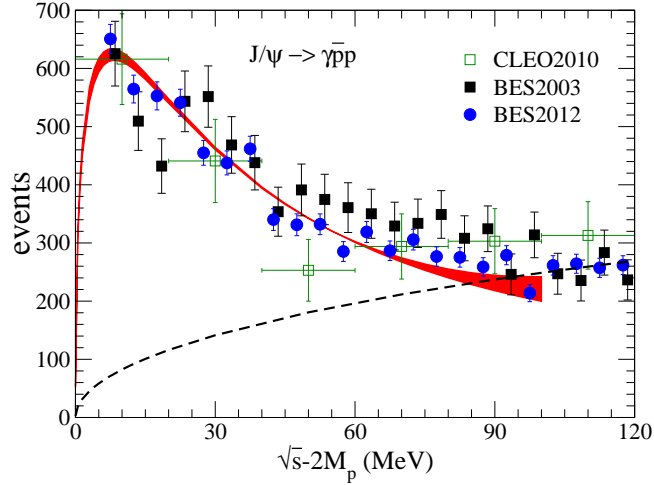} 
\caption{$\bar pp$ spectrum for the decay $J/\psi\to\gamma\bar pp$.
The band represents our result. The dashed curve denotes the phase space behavior.
Data are taken from Refs.~\cite{BES2003,BES2012,CLEOpsip}.
The measurement of Ref.~\cite{BES2012} is adopted for the scale.
The data for the BES measurement from 2003 have been shifted 1\,MeV to the right to discriminate from the new measurement.}
\label{fig:Jpsigamma}
\end{figure}

\begin{figure}
\centering
\includegraphics[scale=0.5,clip=true]{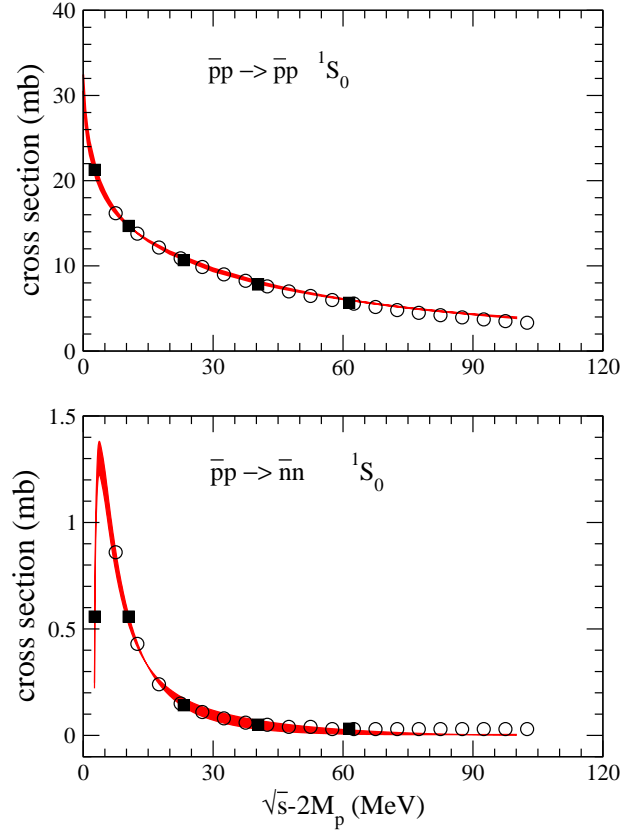}    
\vglue 1ex 
\caption{The $^1S_0$ partial-wave cross sections as a function of the excess energy.
The squares represent the results for the published NNLO potential \cite{NNbar} with
the cutoff combination \{450\,\text{MeV}, 500\,\text{MeV}\}.
The circles indicate the cross sections for the partial-wave amplitudes of Ref.~\cite{Zhou}.
The bands show the results based on the refitted isospin-$1$ $^1S_0$ amplitudes.
}
\label{fig:pwcs}
\end{figure}

After the discovery of the strong threshold enhancement observed in $J/\psi\to\gamma\bar pp$ by BES collaboration \cite{BES2003},
 many explanations have been proposed. Due to its proximity to $\bar pp$ threshold, it is 
speculated to be a $\bar pp$ bound state, or at least, has much to do with $\bar pp$ interaction. 
To take into account the $\bar pp$ FSI, we write the total amplitude symbolically as 
\begin{eqnarray} \label{eq:FSI} 
A=A_0+A_0 G T_{\bar pp},
\end{eqnarray}
where $A_0$ is the elementary production amplitude without considering $\bar pp$ FSI, and in the
second term the off-shell form for $A_0$ is needed since it appears in the integral; $G$ is the free $\bar pp$ Green's function; 
the $T$-matrix elements, $T_{\bar pp}$, can be calculated from Sec.~\ref{sec:NNbar}. 
Writing Eq.~\eqref{eq:FSI} in a partial wave more explicitly, one gets (here for $S-$wave)
\begin{equation}\label{eq:dwba}
A_L=A^0_L \left[1+ \int_0^\infty \frac{dq q^2}{(2\pi)^3} 
\frac{1}{2E_k-2E_q+i0^+}T_{L}(q,k;E_k)\right].
\end{equation}
At low-energy region, assuming $A_0$ has
only a weak energy dependence, one may reasonably approximate as a constant. In this way, in the model there
will be only one parameter, i.e., the overall normalization constant. 
For the observable of event distribution, we are doing a parameter-free calculation, as a matter of fact, in
viewpoint of only the energy dependence.
We stress that the energy dependence of this whole system solely comes from $\bar pp$ interaction since
$A_0$ is treated as a constant. This point indeed verifies the momentous role of $\bar pp$ FSI. In the 
channel $e^+e^-\to\bar pp$, see Sec.~\ref{sec:eeppbar}, the overall constant is used to match the magnitude
of the cross section.
In the process $J/\psi\to\gamma\bar pp$, the lowest allowed quantum number for $\bar pp$ is 
$^1S_0$, and both isospin-0 and 1 are allowed. The result based on the original $^1S_0$ potential
constrained by PWA of Ref.~\cite{Zhou} does not reproduce such a strong enhanced peak near $\bar pp$
threshold. Instead, we perform a combined analysis of the $\bar pp$
scattering data as well as the prominent peak shown by BES data \cite{BES2012}. $T$-matrix is taken
as $T=(T^0+T^1)/2$, where the superscript denotes isospin. The LECs for isospin-0 $^1S_0$ is kept as
what comes from fitting to the PWA of Ref.~\cite{Zhou}, supported by the milder energy dependence of $J/\psi\to\omega \bar pp$ \cite{Jpsiomega}, 
while the four LECs in isospin-1 $^1S_0$
is refitted. So in total, there are 5 free parameters (4 LECs + 1 overall constant). The results are
presented in Fig.~\ref{fig:Jpsigamma} and Fig.~\ref{fig:pwcs}. One can see that the peak around $\bar pp$
threshold is nicely reproduced, and simultaneously, the $^1S_0$ partial-wave cross section of the PWA
as well the original one constructed by us but fitted to PWA are very well reproduced. Then the total 
cross section from our such potential (only $^{31}S_0$ part changes and others do not alter compared to
Ref.~\cite{NNbar}) is thus expected to be in a good agreement with the one calculated from PWA. The protonium
level shift and widths are also examined, and they are also within experimental error bars. It turns that
in order to describe a such prominent peak, we need a $\bar pp$ bound state in isospin-1 $^1S_0$. However, the data above threshold 
is believed to be not capable of pinning down the binding energy and width of this bound state. More 
information on the invariant mass spectrum below $\bar pp$ threshold is needed. In Ref.~\cite{Jpsi}, the
systematic description of the $\bar pp$ mass spectrum in other $J/\psi$ and $\psi'$ decays are achieved.  
In our calculation, the mass difference of proton and neutron, as well as the Coulomb interaction is not considered, and further
work is ongoing.

\subsection{Low-energy $e^+e^-\to \bar pp$ observables }\label{sec:eeppbar}
\begin{figure}
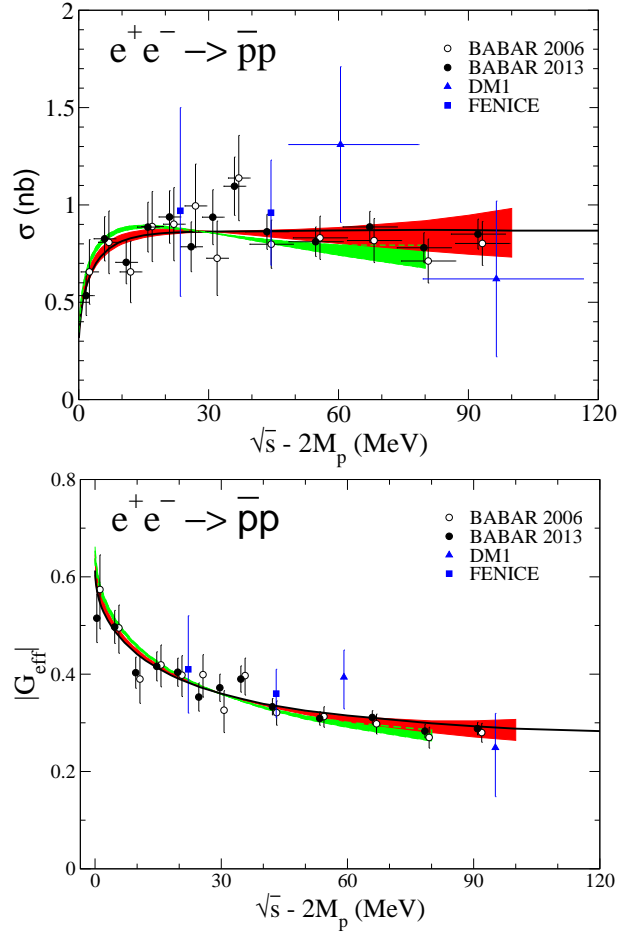

\centering
\includegraphics[scale=0.32,clip=true]{eeppbar.eps}
\includegraphics[scale=0.32,clip=true]{protonFF.eps}
\caption{Cross section and effective form factor of the reaction $e^+e^- \to \bar pp$
as a function of the excess energy.
The data are from the DM1~\cite{DM1} (triangles), FENICE~\cite{FENICE}
(squares), and BaBar~\cite{BaBar2006} (empty circles), \cite{BaBar2013} (filled circles) collaborations.
The red/dark band shows results based on the $\overline NN$ amplitude of the chiral EFT
interaction \cite{NNbar} up to NNLO while the green/light band are those for NLO.
The solid line is the result for the $\overline NN$ amplitude predicted by the
J\"ulich model A(OBE) \cite{Hippchen}.
The BaBar 2006 data are shifted to higher excess energy by 1\,MeV.}
\label{fig:EMFF}
\end{figure}

\begin{figure}
\centering
\vglue 0.6cm
\includegraphics[scale=0.35,clip=ture]{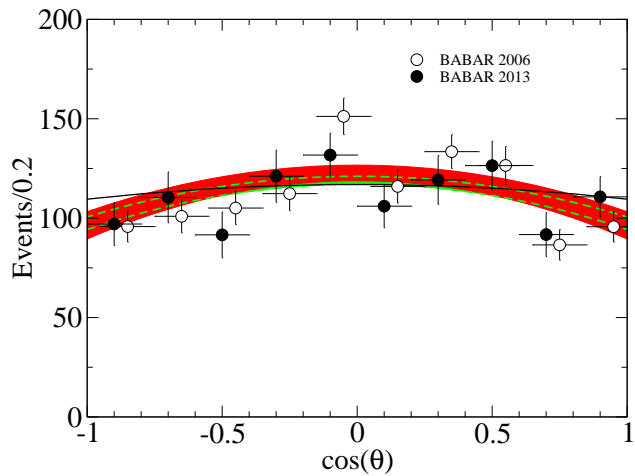}
\caption{Differential cross section for $e^+e^-\to \bar pp$ at
the excess energies of 36.5\,MeV.
The data are an average over [0, 73]\,MeV and are taken from Refs.~\cite{BaBar2006,BaBar2013}.
Same notations as in Fig.~4.}
\label{fig:diff}
\end{figure}

Examining electromagnetic form factors (EMFF) of the proton ($G_E$ and $G_M$) is an efficient way to probe the nucleon structure. 
The reaction $e^+e^-\to \bar pp$, and its inverse one $\bar pp\to e^+e^-$ (these two are related to each other by time reversal operation) are used to measure EMFF. 
The experiment shows a strong energy dependence in proton EMFF close to $\bar pp$ threshold. Recent measurements were done in Refs.~\cite{BaBar2006,BaBar2013}. 
As shown above, at such energy region, the $\bar pp$ FSI plays an important role. And here in the $e^+e^-\to \bar pp$ decay, it is no exception. 
Taking into the fact that one photon exchange should dominate in the decay $e^+e^-\to \bar pp$, and 
thus the only allowed partial wave is the coupled $^3S_1 - {}^3D_1$. This provides an opportunity 
to make a (somewhat) clean prediction, compared to other decay channels where many partial waves are possible and maybe have a comparable significance.  
Including the coupled partial wave $^3S_1 - {}^3D_1$, one could extend Eq.~\eqref{eq:FSI} to a $2 \times 2$ matrix form
\begin{eqnarray}
A_0=\begin{pmatrix} A_0^{SS} & A_0^{SD}\\ A_0^{DS} & A_0^{DD}\end{pmatrix}\,,\quad 
T_{\bar pp}=\begin{pmatrix} T_{\bar pp}^{SS} & T_{\bar pp}^{SD}\\ T_{\bar pp}^{DS} &T_{\bar pp}^{DD}\end{pmatrix}
\end{eqnarray}
The matrix $A_0$, again as before, is the bare production amplitude without $\bar pp$ FSI, and is connected to the bare EMFF $G_E^0$ and $G_M^0$. At near-threshold 
region, they can be approximated as constant. And imposing the condition $G_E=G_M$ we have only one overall normalization constant. Concerning the $\bar NN$ potential
used in this work, both the chiral potential constructed in Ref.~\cite{NNbar} and J\"ulich model A(OBE) \cite{Hippchen} are considered.
The results are shown in Fig.~\ref{fig:EMFF} for the cross section and effective form factor, where for the cross section we fitted to 60\,MeV, and the effective form factor
are calculated from the fitted overall constant. The ratio and the phase difference between the proton form factors $G_E$ and $G_M$ 
are also presented up to the same energy region as the cross section, see Ref.~\cite{eeppbar}.
As can be seen, it reproduces the data rather well. We also calculate the differential cross section at a lower excess
energy of 36.5\,MeV, and the data is nicely reproduced, see Fig.~\ref{fig:diff}. These altogether provide a good description of low-energy data on $e^+e^-\to\bar pp$, and thus
strongly support our speculations of large $\bar pp$ FSI. In the reactions $e^+e^-\to$multipions, $\bar pp$ is also shown to play an important role in the region 
[1750, 1950]\,MeV \cite{eepions}.

\section{Summary}
In summary, we have constructed and established a $\overline NN$ potential in chiral EFT. The resulting phase shifts and inelasticities agree with the partial-wave analysis 
reported in Ref.~\cite{Zhou}. Scattering lengths and the level shift and widths of antiprotonic hydrogen are calculated, and they are in a good agreement with
the experimental information within error bars. With such a $\overline NN$ interaction at hand, we explored the $\bar pp$ FSI in several reactions. For $J/\psi\to\gamma \bar pp$
we perform a combined analysis of experimental events distribution and the $\bar pp$ partial-wave cross section ($^1S_0$ case). In order to describe the prominent peak shown
in $J/\psi\to\gamma \bar pp$, we need a bound state in $^{31}S_0$. But the binding energy can not be well determined. The large $\bar pp$ FSI is also verified in $e^+e^-\to\bar pp$.
The cross section and effective form factor up to the excess energy of 100\,MeV are nicely reproduced. In all these processes, in fact, the whole energy dependences with the $\bar pp$
invariant mass spectrum come solely from the $\bar pp$ final-state interaction, since the bare production amplitude is approximated as constant without energy dependence.

\section*{Acknowledgement}
We thank the organizers for providing the opportunity to present this work. Careful readings of Johann Haidenbauer and Ulf-G.Mei{\ss}ner are also specially acknowledged.
This work is supported in part by the DFG and the NSFC through
funds provided to the Sino-German CRC 110 ``Symmetries and
the Emergence of Structure in QCD''.

\end{document}